\documentstyle[12pt]{article}
\begin{document}

\def\QATOPD#1#2#3#4{{#3 \atopwithdelims#1#2 #4}}
\def\stackunder#1#2{\mathrel{\mathop{#2}\limits_{#1}}}
\def\bea{\begin{eqnarray}}
\def\eea{\end{eqnarray}}
\def\nn{\nonumber}
\def\baselinestretch{1.5}
\def\beq{\begin{equation}}
\def\eeq{\end{equation}}
\def\ba{\beq\new\begin{array}{c}}
\def\ea{\end{array}\eeq}
\def\be{\ba}
\def\ee{\ea}
\def\stackreb#1#2{\mathrel{\mathop{#2}\limits_{#1}}}
\def\Tr{{\rm Tr}}
\def\res{{\rm res}}
\def\f{1\over}
\parskip=0.4em
\makeatletter
\newdimen\normalarrayskip              
\newdimen\minarrayskip                 
\normalarrayskip\baselineskip
\minarrayskip\jot
\newif\ifold             \oldtrue            \def\new{\oldfalse}
\def\arraymode{\ifold\relax\else\displaystyle\fi} 
\def\eqnumphantom{\phantom{(\theequation)}}     
\def\@arrayskip{\ifold\baselineskip\z@\lineskip\z@
     \else
     \baselineskip\minarrayskip\lineskip2\minarrayskip\fi}
\def\@arrayclassz{\ifcase \@lastchclass \@acolampacol \or
\@ampacol \or \or \or \@addamp \or
   \@acolampacol \or \@firstampfalse \@acol \fi
\edef\@preamble{\@preamble
  \ifcase \@chnum
     \hfil$\relax\arraymode\@sharp$\hfil
     \or $\relax\arraymode\@sharp$\hfil
     \or \hfil$\relax\arraymode\@sharp$\fi}}
\def\@array[#1]#2{\setbox\@arstrutbox=\hbox{\vrule
     height\arraystretch \ht\strutbox
     depth\arraystretch \dp\strutbox
     width\z@}\@mkpream{#2}\edef\@preamble{\halign
\noexpand\@halignto
\bgroup \tabskip\z@ \@arstrut \@preamble \tabskip\z@ \cr}%
\let\@startpbox\@@startpbox \let\@endpbox\@@endpbox
  \if #1t\vtop \else \if#1b\vbox \else \vcenter \fi\fi
  \bgroup \let\par\relax
  \let\@sharp##\let\protect\relax
  \@arrayskip\@preamble}
%
%
%
%
\def\eqnarray{\stepcounter{equation}%
              \let\@currentlabel=\theequation
              \global\@eqnswtrue
              \global\@eqcnt\z@
              \tabskip\@centering
              \let\\=\@eqncr
              $$%
 \halign to \displaywidth\bgroup
    \eqnumphantom\@eqnsel\hskip\@centering
    $\displaystyle \tabskip\z@ {##}$%
    \global\@eqcnt\@ne \hskip 2\arraycolsep
         $\displaystyle\arraymode{##}$\hfil
    \global\@eqcnt\tw@ \hskip 2\arraycolsep
         $\displaystyle\tabskip\z@{##}$\hfil
         \tabskip\@centering
    &{##}\tabskip\z@\cr}


\setcounter{footnote}0
\begin{center}
\hfill ITEP/TH-42/98\\
\hfill{ BUTP/23-98}\\
\vspace{0.3in}
{\LARGE\bf Towards the low energy effective degrees of freedom
in N=2 SUSY theories: branes and integrability}
\date{today}

\bigskip {\Large A.Gorsky \footnote{
permanent address: ITEP, Moscow, 117259, B.Cheryomushkinskaya 25;
Talk given at ``Quarks-98'', Suzdal, May 98}}
\\
\bigskip
Institute for Theoretical Physics, University of Bern\\

\end{center}
\bigskip

\begin{abstract}
We discuss the brane interpretation of the integrable dynamics behind the
exact solution to the N=2 SUSY YM theory. Degrees of freedom
in the first integrable system
responsible for the  spectral Riemann
surfaces comes from the hidden Higgs branch of the moduli space.  The
second integrable system of the Whitham type yields the
dynamics on the Coulomb branch and can be considered as the  scattering of
branes.

\end{abstract}


The description of the strong coupling regime in
the quantum field theory remains a
challenging problem and the main hope is connected to discovery of 
new proper degrees
of freedom which would provide the perturbative expansion
distinct from the initial
one. The first sucsessful derivation of the low energy
effective action in N=2
SUSY Yang-Mills theory clearly shows that solution of the theory involves new
ingredients which are not familiar
in this context before like
Riemann surfaces and meromorphic differentials on it \cite{sw}.

The general structure of the effective actions
is defined by the symmetry arguments,
in particular they should respect the Ward
identies coming from the bare field theory.
For example the chiral symmetry  fixes the chiral
Lagrangian in QCD and the conformal symmetry  provides 
the dilaton effective
actions in N=0 and N=1 YM theories. Since the effective actions have a symmetry
origin one can expect universality properties and generically different
UV theories can flow to the same IR ones.
It is the symmetry origin of the effective
actions that leads to  the appearence
of the integrable systems on the scene. The point is that the phase spaces
for the integrable systems coincide with some moduli space or the cotangent
bundle to the moduli space. We can mention  KdV hierarchy related to the
moduli of the complex structures of the Riemann surfaces, the Toda lattice
related to the moduli of the flat connections or Hitchin like systems
connected with the moduli of the holomorphic vector bundles. In any case
moduli spaces come from some additional symmetry of the problem.

On the other hand to get the effective action one has to integrate
over the moduli spaces of the nonperturbative configurations in the theory. 
Nonperturbative configurations relevant in different dimensions
are instantons, monopoles, vortexes or solitons. If one restricts himself to
the 4d theories all essential moduli spaces like moduli of flat connections
or monopoles can be derived by the reduction proceedure from the universal
instanton moduli space. In terms of the integrable systems the problem
of  calculation of  contributions from the nonperturbative fluctuations
to the effective action can be reformulated as a calculation of some
expectation values  in the integrable systems on the moduli
space of these fluctuations.

Identification of the variables in the integrable system responsible for
some effective action is a complicated problem. At a moment there is no
universal way to introduce the proper variables in the theories which
are not topological ones but there is some experience in 2d theories
\cite{vafacec} which suggests to identify the nonperturbative
transition amplitudes among the vacuum states
as the dynamical variables. As for the "space-time" variables, coupling
constants and sources are the most promising candidates. It is
expected that the partition function evaluated in the low-energy effective
theory is the so called $\tau$ - function in integrable hierarchy which
is the generating function for the conserved integrals of motion.
The particular solution of the equations of motion in the dynamical system
is selected by applying the Ward identities to the partition function of
the effective theory.

The arguments above explain the reason for the search of some integrable
structures behind the Seiberg-Witten solution to N=2 SUSY Yang-Mills
theory. This integrable structures which capture the hidden symmetry structure
have been found in \cite{gkmmm} where it was shown that $A_{N_{c}}$
affine Toda chain governs the low energy effective action and BPS
spectrum of pure N=2 SYM theory. The generalization to the theories
with matter involves Calogero-Moser integrable system for the adjoint
matter \cite{dw}  and XXX spin chain for the fundamental matter \cite{chains}. 
When generalizing
to 5d relativistic Toda chain appears to be relevant for the pure
gauge theory \cite{nikita} while anizotropic XXZ chain for SQCD \cite{ggm2}. 
At the next step
completely anisotropic XYZ chain has been suggested as a guide for 
6d SQCD \cite{ggm2}
while the generalization to the group product case is described by
the higher spin magnets \cite{ggm1}. Therefore 
there are no doubts in the validity
of the mapping between effective low-energy efective theories and
integrable finite dimensional systems.

The list of correspondences between two seemingly different issues
looks as follows. The solution of the classical equation of motion
in the integrable system can be expressed in terms of higher henus
Riemann surface which can be mapped to the complex Liouville
tori of the dynamical system. It is this Riemann surface enters
Seiberg-Witten solution, and the meromorphic differential introduced
to formulate the solution coincides with the action differential
in the dynamical system in the separated variables. The Coulomb
moduli space in N=2 theories is identified with the space of the
integrals of motion in the dynamical system, for example 
$Tr{\phi}^{2}$ where $\phi$ is the adjoint scalar field coincides
with  the
Hamiltonian for the periodical Toda system. The parameters of
the field theory like masses or $\Lambda_{QCD}$
determine the parameters and couplings in the integrable system.
For instance in SQCD fundamental masses provide the local Casimirs in the
periodical spin chains. The full list of interrelations and references
can be found in the review \cite{mars}.

In spite of a lot of supporting facts it is necessary to get more
transparent explanation of the origin of integrability in this
context. To this aim let us discuss the moduli spaces in the problem
at hands. Classically there is only Coulomb branch of the moduli
space in pure gauge theory so one can expect dynamical system associated
with such phase space. Coulomb branch can be considered as a special
Kahler manifold  \cite{sw} while the Hitchin like dynamical system responsible
for the model has a hyperKahler phase space
\cite{hitchin}. The resolution of the
contradiction comes from the hidden Higgs-like branch which has
purely nonperturbative nature
\cite{gorbr,ggm1}. It is the  dynamical system on this hidden
phase space provides the integrable system of the Hitchin 
or spin chain type. Therefore
there are two moduli spaces in our problem and one expects a pair
of dynamical systems. This is what we have indeed; dynamical system
on the Higgs branch yields the Hitchin like dynamics with the 
associated Riemann surfaces while the integrable system on the 
Coulomb branch gives rise to the Whitham dynamics. The ``physical''
meaning of the Hitchin system is to incorporate the nonperturbative
instanton like contributions to the effective action in the supersymmetric
way while the Whitham dynamics is nothing 
but the RG flows in the model \cite{gkmmm}.

The next evident question is about the degrees of freedom in 
both dynamical systems. The claim is that all degrees of the freedom can
be identified with the collective coordinates of a particular brane
configuration. First let us explain where the Higgs branch comes
from. The basic illustrative example for the 
derivation of the hyperKahler moduli
space in terms of branes is the description of ADHM data 
as a moduli for a  system  of coupled D1-D5
or D0-D4 branes \cite{doug}. If the gauge fields are 
independent on some  dimension
one derives Nahm description of the monopole moduli space in terms 
of D1-D3 branes configuration \cite{diac}. The transition from ADHM 
data to the Nahm ones can be treated as a T duality transformation.
At the next step the  hyperkahler Hitchin space can be obtained by
reducing the dependence (or additional T duality transformation) 
on one more dimension. This corresponds to  the system of D2 branes
wrapped around some surface $\Sigma$ holomorphically embedded in some
manifold. The most relevant example concerns
$T^{2}$  embedded into K3 manifold \cite{vafatop}. The T duality along
the torus transforms it to the system of D0 branes on the dual
torus, which is the most close picture for the Toda dynamics in terms
of D0 branes. The related discussion for the derivation of the 
Hitchin spaces in terms of instantons 
on $R^{2}\times T^{2}$ can be found in \cite{kap}.

Let us now proceed to the explicit brane picture for the
N=2 theories. There are different ways to get it, one involves
10d string theory which compactified on the manifold containing
the Toda chain spectral curve \cite{vafa}, or the M theory with 
M5 brane wrapped around the noncompact surface which can be
obtained from the spectral curve by deleting the finite 
number of points \cite{wittenm}. This picture can be considered as the 
perturbative one  and nonperturbative degrees of freedom have to be
added. For this purpose it is useful to consider IIA projection of
the M theory which involves $N_{c}$ D4 branes between two NS5 branes
located on a distance $\frac{l_{s}}{g^{2}}$ along, say $x_{6}$ 
direction. Field theory is defined on D4 branes worldvolume \cite{witbr} and
the extensive review cocerning 
the derivation of the field theories from branes 
can be found in \cite{givkut}. The additional ingredient yielding
the hidden Higgs branch comes from the set of $N_{c}$ D0 branes,
one per each D4 brane \cite{gorbr,ggm1}. It is 
known that D0 on D4 brane behaves 
as a abelian point-like instanton but now we have the system of
interacting D0 branes. The coupling constant is provided by the
$\Lambda_{QCD}$ parameter which can be most naturally obtained
from the mass of the adjoint scalar breaking N=4 to N=2 via
dimensional transmutation proceedure.

One way to explain the need for the additional D0,s
in IIA theory or KK modes in M theory looks as follows.
It is known that any finite-dimensional integrable system
with the spectral parameter allows the canonical transformation
to the variables -- spectral
curve with the linear bundle.
The spectral curve place is transparent and KK modes provide
the linear bundle. As we have already noted they are responsible
for the nonperturbative contribution but the summation of the 
infinite instanton sums into the finite number degrees of freedom
remains the challenging problem. 
It is worth noting that both canonical coordinates in the dynamical
system comes from the coordinates of D0 branes in different
dimensions.
The necessity for the additional
nonperturbative degrees of freedom has been also discussed in
\cite{do}.

To show how the objects familiar in the integrability world translate
into the brane language consider two examples.
First let us consider the equations of motion in the Toda chain
which has the Lax form
\beq
\frac{dT}{ds}=[T,A]
\eeq
with some $N_{c}\times N_{c}$ matrixes T and A. The Lax matrix T 
can be related to Nahm matrix for the chain of monopoles using 
the identifications of the spectral curves for cyclic monopole
configuration and periodic Toda chain \cite{sut}.
All these
results in the following expression
for the Toda Lax operator in terms of the Nahm matrixes $T_{i}$
\be
T=T_{1}+iT_{2}-2iT_{3}{\rho}+(T_{1}-iT_{2}){\rho}^{2}
\ee
\be
T_{1}=\frac{i}{2}\sum_{j=1}q_{j}(E_{+j}+E_{-j})  \\
T_{2}=-\sum_{j=1}q_{j}(E_{+j}-E_{-j})             \\
T_{3}=\frac{i}{2}\sum_{j}p_{j}H_{j} ,
\ee
where E and H are the standard SU(N) generators, $p_{i},q_{i}$ represent
the Toda phase space, and $\rho$ is the coordinate on the $CP^{1}$ above.
This $CP^{1}$ is involved in the twistor construction for
monopoles and a point on $CP^{1}$  defines the complex structure on the
monopole moduli space.
With these definitions Toda equation of motion and Nahm
equation acquire the simple form
\be
\frac{dT}{dt}=[T,A]
\ee
with fixed A.
Having in mind the brane interpretation of the
Nahm data \cite{diac} we can claim that the equations of motion provide 
the conditions for the required supersymmetry of the whole configuration.

Given the dynamical system let us discuss the interpretation 
of the BPS spectrum
in the integrable terms \cite{gorpei}. There are many different 
brane realizations
of the BPS spectrum connected by dualities but the ``integrable'' one
can be described in terms of Lax fermions $\Psi(\lambda)$ - eigenfunctions 
of the Lax
operator
\beq
T\Psi=\lambda\Psi,
\eeq
where $\lambda$ is the spectral parameter in the dynamical system and
simultaneously plays the role of the energy of the spectral fermions.
Toda chain spectral curve plays the role of the solution to the equation
of motion and simultaneously the dispersion law for the Lax fermions.
Therefore it can be shown \cite{gorpei} that the BPS states correspond
to the completely filled forbidden or allowed band for the Lax fermion .

As another example of the validity of the brane-integrability
correspondence mention the possibility to incorporate the fundamental
matter in the gauge theory via branes in two ways. The first one 
concerns the semiinfinite D4 branes while the second one the
set of $N_{f}$ D6 branes. One can
expect two different integrable systems behind and they were found
in \cite{kriph} and \cite{chains}. It was shown in \cite{ggm1} that 
they perfectly correspond to the brane pictures  and it appears that the
equivalence of two representations agrees with some duality property in
the dynamical system. To conclude the discussion of the first
dynamical system let us mention that one can inverse the logic and
use the possible integrable deformations of the dynamical system 
to construct their field theory counterparts. Along this line of
reasoning we can expect some unusual field theories with the several
$\Lambda$  type scales \cite{ggm1}.

Let us proceed now to the second integrable system on the Coulumb branch
of the Whitham type. Whitham dynamics provides
in a most natural way prepotential $\cal{F}$ which yields the low energy
effective action in Seiberg-Witten solution. The prepotential as
the solution of Whitham equations  is related to the action 
calculated on the solution with $\Lambda_{QCD}$ playing the role of
the time variable. It is illustrative to consider the identity \cite{ma}
\beq 
\frac{\partial \cal{F}}{\partial log\Lambda}=\beta<Tr\phi^{2}>
\eeq
as a relation between action and Hamiltonian, which simultaneously
can be treated as the superconformal Ward identity.
Remarkably, the prepotential is closely related to the topological 4d
theories which can be supported by investigation of the
WDVV like equations in 4d \cite{mmm}. Recently there was some progress
concerning the topological properties of N=2 theories which can
be formulated in terms of Donaldson theory for the instanton
moduli spaces \cite{donal,lns}. It appears
that this approach is consistent with the Whitham flows in the
nonlinear approximation \cite{whithdon} and the second derivative
of the prepotential which can be considered as the correlator of the
proper operators in N=2 pure gauge theory can be equally calculated
within Donaldson  \cite{lns} and Whitham setups

\be
\frac{\partial^2{\cal F}}{\partial T^m\partial T^n}
= -\frac{\beta}{2\pi i} \left({\cal H}_{m+1,n+1}
+ \frac{\beta}{mn}\frac{\partial {\cal H}_{m+1}}{\partial a^i}
\frac{\partial {\cal H}_{n+1}}{\partial a^j}
\partial^2_{ij} \log \theta_E(\vec 0|{\cal T})\right)
\label{2der}
\ee
In these formulas the gauge group is $G = SU(N)$,
parameter $\beta = 2N$, $m,n = 1,\ldots,N-1$.
$T_{n}$ are the Whitham times and variables $a^{i}$ are the 
standard Seiberg-Witten integrals of the meromorphic differential
over the i-th cycle on the spectral curve.
${\cal H}_{m,n}$ are certain homogeneous combinations of
$h_k$, defined in terms of the $h$-dependent polynomial
$P(\lambda)$ which is used to describe the Seiberg-Witten
(Toda-chain) spectral curves:
\be
{\cal H}_{m+1,n+1} =
-\frac{N}{mn}
{\res}_\infty\left(P^{n/N}(\lambda)d P^{m/N}_+(\lambda)\right)
= {\cal H}_{n+1,m+1}
\ee
and
\be
{\cal H}_{n+1} \equiv {\cal H}_{n+1,2}
= -\frac{N}{n}{\res}_\infty P^{n/N}(\lambda)
d\lambda  = h_{n+1} + O(h^2).
\ee
 
To recognize the brane realization of the 
second integrable system of the Whitham type let us adopt
slightly different perspective
from the F-theory on the elliptically fibered K3 which is equivalent
to the orientifold of type IIB theory or, after T duality, to type I theory
on $T^{2}$.
Due to \cite{sen} we can treate the N=2 d=4 theory as a world volume theory
of 3-branes in the background of the splitted orientifold  planes placed at
points $\pm\Lambda^{2}$ in the $u=Tr\phi^{2}$ complex plane
for the SU(2) case. We assume that the possible masses
of the fundamental matter tend to infinity so we are in the pure YM case.

Now we have to consider the dynamics of the
3-branes
in the directions transverse to the background 7 branes.
The arising dynamics is very
transparent  in the SU(2) case. Let us recall that Whitham dynamics
for SU(2) case is governed by the solution of the first Gurevich-Pitaevskii
problem \cite{gkmmm}
which can be easily interpreted as follows. At the initial moment of
evolution 3-branes coincide with one of the orientifold planes and with the
another planes at the end of the evolution.
To analyze the Whitham dynamics in SU(2) case it is convenient to use the
following form of the spectral curve
\be
y^{2}=(x^{2}-\Lambda^{4})(x-u).
\ee
The point u represents the position of two 3-branes (which are at
$\pm \sqrt u$
in the $\phi$ plane) and  another branching points  give the fixed
positions of the background branes. The branching point u moves according to
the Whitham dynamics for the one-gap KdV solution which 
corresponds to the Seiberg-Witten solution.  
\be
\eta(x,t)=2dn^{2}[\frac{1}{\sqrt6}(x-\frac{1+s^{2}}{3}t,s)]-(1-s^{2})
\ee
where
\be
\frac{1+s^{2}}{3}
-\frac{2s^{2}(1-s^{2})K(s)}{3(E(s)-(1-s^{2})K(s))}=\frac{x}{t},
\ee
K(s) and E(s) are the elliptic moduli and
$s^{2}=\frac{u+\Lambda^{2}}{2\Lambda^{2}}$.
In terms of the automodel variable $\theta=\frac{x}{t}$
the left background brane corresponds to $\theta=-1$ while the right to
$\theta=\frac{2}{3}$. Quasiclassical tau-function of this solution provides
the prepotential for the SU(2) theory $\cal{F}$=log ${\tau}_{qcl}$.
There is no the analogous simple 
brane picture for the Whitham dynamics for
the higher rank groups at a moment.

Therefore we have presented interpretation of the pair of the
integrable dynamical systems in the brane terms. It is clear that 
there are a lot of open questions concerning the integrable structures
behind the nonperturbative SUSY YM dynamics. We can mention for  instance
the spectrum generating algebras in the integrable systems which
are expected to be related to the Nakajima,s algebras on the homologies
of the instanton moduli spaces, the embedding of the finite dimensional
systems as the special solution to the integrable field theories
or clarification of the integrable structure behind N=1 theories.

I am indebted to S.Gukov, A.Losev, 
A.Marshakov, A.Mironov, A.Morozov and N.Nekrasov
for the collaboration and interesting discussions.
This work is supported in part by grants INTAS-96-0482, CRDF-RP2-132
and Schweizerischer Nationalfonds.

\end{document}